# Negative Photoconductance in van der Waals Heterostructures-Based Floating Gate Phototransistor

Yu Wang,[†,▲] Erfu Liu,[†,▲], Anyuan Gao,[†] Tianjun Cao,[†] Mingsheng Long,[†] Chen Pan,[†] Lili Zhang,[†] Junwen Zeng,[†] Chenyu Wang,[†] Weida Hu,[‡] Shi-Jun Liang[*†] and Feng Miao[*†]

[†] National Laboratory of Solid State Microstructures, School of Physics, Collaborative Innovation Center of Advanced Microstructures, Nanjing University, Nanjing 210093, China.

[‡] National Laboratory for Infrared Physics, Shanghai Institute of Technical Physics, Chinese Academy of Sciences, Shanghai 200083, China.

**ABSTRACT:** Van der Waals (vdW) heterostructures made of two-dimensional materials have been demonstrated to be versatile architectures for optoelectronic applications due to strong light-matter interactions. However, most of light-controlled phenomena and applications in the vdW heterostructures rely on positive photoconductance (PPC). Negative photoconductance (NPC) has not yet been reported in vdW heterostructures. Here we report the observation of the NPC in $ReS_2$/h-BN/$MoS_2$ vdW heterostructures-based floating gate phototransistor. The fabricated devices exhibit excellent performance of nonvolatile memory without light illumination. More interestingly, we observe a gate-tunable transition between the PPC and the NPC under the light illumination. The observed NPC phenomenon can be attributed to the charge transfer between floating gate and conduction channel. Furthermore, we show that the control of NPC through light intensity is promising in realization of light-tunable multi-bit memory devices. Our results may enable potential applications in multifunctional memories and optoelectronic devices.

Exposure of semiconductors to light with energy larger than band gap generates excess mobile charges (electrons or holes) in the conduction and/or valence bands, and leads to an increase in electrical conductivity of the semiconductors, which is referred to positive photoconductance (PPC). The effect is of crucial importance to many optoelectronics applications, such as photodetectors,[1,2] photoelectronic memory.[3] Conversely, the negative photoconductance (NPC) refers to a reduced conductivity of semiconductors under light illumination. With distinctive mechanisms, the NPC phenomenon has been reported in the bulk semiconductors[4,5] and low-dimensional structures.[6-9] Moreover, it has been demonstrated that the NPC holds promise in constructing optoelectronic devices with low power consumption and high-speed frequency response.[10,11] Combining the PPC and the NPC in one device is expected to extend the functions of conventional optoelectronic devices and may enable potential applications in optoelectronic devices.

Van der Waals (vdW) heterostructures created by stacking different two-dimensional (2D) materials with highly distinct electronic states exhibit intriguing optical[12-14] and electronic properties.[15-17] Employing vdW heterostructures integrating diverse layers of metal and semiconductor or insulator holds tremendous promise for revealing many intriguing physical phenomena and designing electronic devices with extended functionalities and superior performance.[18-23] Two types of intriguing applications based on vdW heterostructures are gate-tunable optoelectronic devices[24-29] and photo-controllable memory devices.[30-34] Most of these applications are based on the PPC mechanism. However, the NPC has not yet been reported in vdW heterostructures.

In this work, we observe the NPC phenomenon in $ReS_2$/h-BN/$MoS_2$ vdW heterostructures-based floating gate phototransistor. The fabricated floating-gate architecture devices show excellent memory properties, such as high on/off current ratio (over $10^7$), good retention (more than 10000s) and endurance (over 2000 cycles), More importantly, the transition between the PPC and the NPC is gate-tunable in the floating gate phototransistor. With these significant properties, the $ReS_2$/h-BN/$MoS_2$ vdW heterostructures may provide a different avenue for the development of in light-controlled nonvolatile memory devices and multifunctional photoconductive devices.

**Results/Discussion**

To fabricate vdW heterostructure floating gate devices, few-layer MoS$_2$ flakes (as floating gate) were firstly exfoliated onto the SiO$_2$/Si substrate. Then h-BN flakes (as potential barrier layer) and ReS$_2$ flakes (as conduction channel) were transferred onto the top of MoS$_2$ flakes. The devices fabrication details are given in the section of **Methods**. The experimental setup of ReS$_2$/h-BN/MoS$_2$ vdW heterostructure floating gate devices and its equivalent circuit diagram are shown in Figure 1a and Figure 1b, respectively. $V_{ds}$=0.1 V is the source/drain bias voltage and $V_{cg}$ represents the control gate voltage. Figure 1c shows the optical image of a typical vdW heterostructure floating gate device. Raman signatures of MoS$_2$ (E$_{2g}$ peak at 383 cm$^{-1}$ and A$_{1g}$ peak at 408 cm$^{-1}$) and ReS$_2$ (E$_{2g}$ peaks at 163 cm$^{-1}$ and A$_{1g}$ at 213 cm$^{-1}$) indicate successful fabrication of the vdW heterostructure floating gate devices (Figure 1d).

The fabricated ReS$_2$/h-BN/MoS$_2$ vdW heterostructure can be used as a nonvolatile memory device. To characterize its memory performance, we measured transfer characteristics, retention and endurance properties, with results shown in Figure 2. The hysteresis can be observed in the transfer characteristics (Figure 2a) for different scanning ranges of $V_{cg}$. At $V_{cg}$=0 V, the ON/OFF current ratio exceeds 10$^7$, which is larger than that of charge-trapping memory devices in 2D materials[35,36] (*e.g.* black phosphorus and MoS$_2$) and floating-gate vdW heterostructures.[37-39] Gate-tunable hysteresis resulting from charge trapping in the MoS$_2$ floating gate is indicative of the storage capability of memory device. We then define the width of memory window $\Delta V$ as difference between the threshold voltage for low and high resistive states to check the store capability of charge in MoS$_2$, and find that it is linearly proportional to the maximum $V_{cg}$. Sweeping $V_{cg}$ from -50 V to 50 V, we obtain a large width of memory window ($\Delta V \approx$80 V), suggesting the high reliability of the memory device. Similarly, we also observe large memory windows in ReS$_2$/h-BN/Graphene and ReS$_2$/h-BN/WSe$_2$ vdW heterostructure floating gate devices (Figure S1 and S2 in the Supporting Information), which proves that the vdW heterostructure is an ideal platform for constructing nonvolatile memory devices.

The operating mechanism of the fabricated vdW heterostructure as memory device can be readily explained *via* the energy band diagram (Figure 2c) of $MoS_2$/h-BN/$ReS_2$ vdW heterostructure. When $V_{cg}$ is swept towards large positive value, electrons are transferred from $ReS_2$ into $MoS_2$ by tunneling through h-BN (the left panel of Figure 2c), which defines the program state of the memory device. Weakened gating effect induces a positive shift of threshold voltage. If $V_{cg}$ is swept towards large negative value, electrons are driven back from $MoS_2$ to $ReS_2$ *via* tunneling, which defines the erase state. Electrons departing from $MoS_2$ cause a negative shift of the threshold voltage (the right panel of Figure 2c).

The vdW heterostructure floating gate device shows good retention and endurance properties. Applying $V_{cg}$ pulses with 300 ms duration onto the vdW heterostructure floating gate device, the measured currents at program (+30 V) and erase states (-30 V) are very stable over $10^4$ s (Figure 2d). The negligible decay in the measured current suggests little charge leakage from $MoS_2$. The fabricated vdW heterostructure floating gate device also demonstrates strong robustness (Figure 2e). With the same pulse duration and the source/drain bias, we observe that the measured current remains unchanged over 2000 cycles (the inset of Figure 2e), which is more robust against $MoS_2$ flakes memory device based on metal nanoparticle floating gate.[31] The excellent performance of the memory devices may be attributed to the strongly-modulated carrier density in $ReS_2$ by floating gate and the charge transfer between $MoS_2$ and $ReS_2$.

The fabricated vdW heterostructure floating gate device acts not only as a memory device, but also a floating-gate phototransistor.[40] $ReS_2$ remains a direct bandgap during thinning from bulk to monolayer.[41,42] Moreover, it has an ultra-high photoresponsivity[43] and anisotropy.[44,45] Utilizing the high photoresponse of $ReS_2$ and the efficient control of carrier density through floating gate may give rise to special phenomena and enhance the performance of memory devices. Figure 3a shows the photoresponse of $ReS_2$/h-BN/$MoS_2$ heterostructure devices to light excitation of 520 nm wavelength at two different voltage pulses (+60 V and -60 V). With the applied negative voltage pulses lasting for 1s, it is immediately removed and the light is simultaneously switched on.

Surprisingly, we find that the measured current (red curve) quickly drops from $10^{-6}$ A down to $10^{-7}$ A, which indicates the occurrence of the negative photoconductance. In contrast, the positive photoconductance (blue curve) occurs for the applied positive voltage pulses with the same duration. This indicates that the transition between the NPC and the PPC is gate-tunable in the vdW heterostructure floating gate device. It is worth noting that the NPC phenomenon has been observed in the 2D materials such as graphene,[46] monolayer $MoS_2$[47] and black phosphorous.[48] All these reported NPC phenomena in 2D materials arise from the reduction in carrier mobility of 2D materials under light illumination. For the NPC observed in the graphene, the $O_2$ molecules and/or OH groups absorbed on the graphene surface as the scattering centers reduce the electrical conductivity under light illumination, giving rise to the NPC. The NPC in the graphene is uncontrollable as it is strongly dependent on the concentration of absorbents. Different from the origin to the NPC observed in the graphene, the NPC in the single-layer $MoS_2$ results from the formation of trions under femtosecond laser excitation. But the short lifetime of trions would make little difference to the current level in optoelectric devices with relative long transport channel. For the NPC occuring in the black phosphorous, the enhanced electron-phonon scattering is the main mechanism due to the low thermal conductivity of substrate. However, aforementioned mechanisms reported in the 2D materials are irrelevant to the NPC occuring in the vdW heterostructure.

The occurrence of the NPC phenomenon in the vdW heterostructure floating gate phototransistor is related to the floating gate structure. To prove that, we measured the photoresponse of $ReS_2$/h-BN heterostructure device without the floating gate for comparison (Figure 3b). A negative gate voltage pulse with 1 s width is applied to the device. As expected, the current (between electrodes 3 and 4) in the $ReS_2$/h-BN heterostructure device increases from 2.9 μA to 3.1 μA under the light illumination of 520 nm wavelength after removing the negative gate voltage pulse (-60 V), which is due to the positive photoconductance (PPC) arising from the intrinsic photoresponse (red curve) of $ReS_2$ in the $ReS_2$/h-BN heterostructure device. In contrast to the current behavior in the $ReS_2$/h-BN heterostructure device, the measured current (between electrodes 1 and 2) in the ReS$_2$/h-BN/MoS$_2$ heterostructure device quickly drops from 3.3 µA down to 0.24 µA in 30 ms (blue curve). Note that the NPC phenomenon can be also observed in the ReS$_2$/h-BN/Graphene and ReS$_2$/h-BN/WSe$_2$ as well as MoS$_2$/h-BN/MoS$_2$ vdW heterostructure floating-gate phototransistors (Figure S1 and Figure S2 as well as Figure S3 in the Supporting Information), indicating that the band alignment is not the critical factor in accounting for NPC phenomenon. Essentially, the floating gate in the vdW heterostructure affects the NPC through the amount of trapped charges controlled by the control gate. To quantitatively analyze the gate voltage dependence of the NPC, we plot the current change $\Delta I_{ds}$ (current difference between after and before light illumination) as a function of $V_{cg}$ (Figure 3c). We find that $\Delta I_{ds}$ increases with decreasing $V_{cg}$ until reaching -30V, where $\Delta I_{ds}$ becomes saturated. If continuing to reduce $V_{cg}$ down to -10 V, the NPC phenomenon almost disappears (Figure S4 in the Supporting Information), suggesting that the NPC is highly controllable *via* the control gate voltage.

The NPC mechanism reported in the vdW heterostructure is completely different from the hot carrier trapping observed in conventional Si [9] and InAs [10] nanowires, and light excitation induced reduction in the carrier mobility of 2D materials (*e.g.* graphene, MoS$_2$ and black phosphorus).[46-49] The origin of the NPC phenomenon observed in the ReS$_2$/h-BN/MoS$_2$ vdW heterostructure can be elucidated by charge transfer between the floating layer and the conduction channel (Figure 3d). The control gate voltage is initially set to -60 V to make holes accumulated in the MoS$_2$. After removing the control gate voltage, holes are trapped in the MoS$_2$ layer and electrons are simultaneously accumulated in the ReS$_2$ conduction channel (the upper panel of Figure 3d). If turning on light, electron population in the conduction band of ReS$_2$ is significantly enhanced by photon-generated electrons. As a result, there is an abrupt increase in the number of electrons transferring from the ReS$_2$ to the MoS$_2$ by Fowler-Nordheim tunneling (the lower panel of Figure 3d). The electrons from the ReS$_2$ would recombine with the holes trapped in the MoS$_2$, thus the control of the floating gate over the ReS$_2$ conduction channel becomes weak. The weakened floating gate effect leads to less electrons accumulated in the ReS$_2$, then the NPC phenomenon occurs. According to experimental results shown in Figure 3a, and 3c, the NPC not only depends on the polarity but also the magnitude of control gate voltage. When the control gate voltage is initially set to a smaller negative value before removing, there are few holes trapped in the MoS$_2$, resulting in absence of the NPC phenomenon. Note that we have excluded the contribution from the light-generated carriers in the MoS$_2$ by a control experiment of using 71.5 nm thick h-BN as potential barrier. The comparison reveals that the vdW heterostructures with too thick h-BN cannot give rise to the NPC within the gate voltage of interest, because the thick h-BN barrier kills electrons tunneling between MoS$_2$ and ReS$_2$ (Figure S5b in the Supporting Information) under a relatively small control gate voltage (*e.g.* - 40 V). However we can observe a weak feature of the NPC when the gate voltage is increased to a much larger value (Figure S5d in the Supporting Information). We further find that 4.4 nm thin h-BN doesn't work well as a potential barrier, and we only observe the intrinsic photoresponse of ReS$_2$ as trapped carriers could escape from the floating-gate layer rapidly (Figure S6 in the Supporting Information). Therefore, we conclude that the thickness of h-BN as potential barrier is a critical factor in determining occurrence of the NPC in the vdW heterostructures. Only the vdW heterostructure devices with appropriate thickness h-BN as potential barrier enable the observation of the NPC within the applied gate voltage of interest.

The NPC phenomenon is also dependent on the photo-generated electron density. Figure 4a shows light-induced current change under various intensities. We observe that the NPC phenomenon occurs when the light intensity varies from 1 to 738 μW. More interestingly, a sharp transition between the NPC and the PPC is observed when the light intensity is larger than 85 μW. Larger light intensity generates more electrons in the ReS$_2$, consequently, more holes trapped in the MoS$_2$ can be recombined. When most of trapped holes in the MoS$_2$ are recombined, subsequent light-generated electrons are unable to tunnel to MoS$_2$ due to weaker electric field across h-BN layer, therefore they will be accumulated in the ReS$_2$ to induce the transition between the NPC and the PPC. Under the light illumination of lower power, a long tail appears in the curves of current vs time, which may be due to the weakened control of MoS$_2$ floating gate over the ReS$_2$ channel. Once light is switched on, light-generated electrons in the ReS$_2$ instantly recombines with some trapped holes in MoS$_2$, leading to an immediate reduction in current. The less trapped holes in the MoS$_2$ make electric field across h-BN barrier gradually become weaker, and only fewer photo-generated electrons can tunnel to MoS$_2$ to recombine with the remaining holes, which accounts for the long tail in the curves of current vs time.

Design of memory device with extended functionality may benefit from the NPC.[50] Recent work has shown that the multi-bit photoelectronic memory can be achieved through the PPC mechanism.[31,32] To demonstrate the potential application of the NPC in the multi-bit photoelectronic memory, we traced the current change at different light intensities (0.3, 3, and 10 μW), with results shown in Figure 4b. The light is first illuminated onto the device for 10 s and then switched off. With the light on, the current starts to decrease similar to that in Figure 4a. When the light is switched off, the current abruptly drops down to a minimum before slowly rising to a saturated current level. Different light intensity gives rise to a distinctive minimum current level, which we measure as the readout current state. We label these different readout current values as states ('1','2','3'). The current level without light illumination as '0' state is defined as well. The readout current as a function of state (Figure 4c) indicates that the NPC is promising in the development of multi-bit memory devices.

**Conclusions**

In summary, we report the experimental observation of the negative photoconductance in the ReS$_2$/h-BN/MoS$_2$ vdW heterostructures phototransistor. More interestingly, we realize a gate-tunable transition between the PPC and the NPC in the vdW heterostructures phototransistor. The observed NPC is attributed to the light-generated electrons tunneling from the floating gate to the conduction channel. Moreover, we demonstrate that the light intensity tunable NPC is promising in the development of light-controllable multi-bit memory device applications based on vdW heterostructures. Our findings may offer a different avenue for design of photoelectronic memory devices and photodetector devices based on van de Waals heterostructures.

## Methods/Experimental Section

**Materials and Devices.** Heterostructures devices were fabricated by stacking mechanically exfoliated $ReS_2$, h-BN and $MoS_2$ flakes onto the $SiO_2$/Si substrate. Polymer vinylalcohol (PVA) coated on the polydimethylsiloxane (PDMS) was used to pick up the $ReS_2$ and h-BN flakes. A conventional electron-beam lithography process was used to fabricate heterostructures devices, together with the standard electron-beam evaporation of metal electrodes (5 nm Ti/40 nm Au).

**Electrical and Photoresponse Measurements**. Electrical and photoresponse measurements were performed by using a Keithley 2636A dual channel digital source-meter. The lasers with wavelengths of 520nm, 637nm, 830nm and 1310nm were focused on the devices using a 20× objective lens. The devices have been put into a metal box to reduce the noise from environment during photoresponse measurement.

## ASSOCIATED CONTENT

The authors declare no competing financial interests.

## Supporting Information

The Supporting Information is available free of charge on the ACS Publications website at DOI: 10.1021/.

The NPC in $ReS_2$/h-BN/Graphene floating gate device, The NPC in $ReS_2$/h-BN/$WSe_2$ floating gate device, NPC and PPC in $MoS_2$/h-BN/$MoS_2$ vdW heterostructure device, Gate tunable NPC phenomenon, The NPC in floating gate device with different thickness h-BN, NPC under different wavelengths of light (PDF)

## AUTHOR INFORMATION

**Corresponding Authors**

*E-mail: miao@nju.edu.cn (F. M.)

*E-mails: sjliang@nju.edu.cn (S.J. L.)

**Author Contributions**

Y. W. and E.L. contributed equally to this work.

**ACKNOWLEDGEMENTS**

This work was supported in part by the National Key Basic Research Program of China (2015CB921600), the National Natural Science Foundation of China (61625402, 61574076, 11474147), the Fundamental Research Funds for the Central Universities, and the Collaborative Innovation Center of Advanced Microstructures.

# Figure captions

**Figure 1. Characterization of ReS$_2$/h-BN/MoS$_2$ heterostructure devices.** (a) Schematic diagram of the device with light illumination. (b) Measurement circuit of the device. The bias voltages ($V_{ds}$) is 0.1V in all the experiments. $V_{cg}$ is the control gate voltage. (c) Optical microscopy image of a typical ReS$_2$/h-BN/MoS$_2$ heterostructure device. The scale bar is 10 μm. (d) Micro Raman spectrums performed on few-layer ReS$_2$ and MoS$_2$ flakes.

**Figure 2 | Memory characteristics of heterostructure devices.** (a) Transfer curves under different control gate voltages. (b) Memory window width $\Delta V$ as a function of maximum control gate voltage $V_{cg,max}$. (c) Band diagrams of ReS$_2$/h-BN/MoS$_2$ device for $V_{cg} > 0$ V and $V_{cg} < 0$ V. (d) Retention of device. The retention is measured by applying a voltage pulse of -30 V (red circle) or +30 V (blue circle). Pulse width is 1 s and $V_{ds} = 100$ mV. (e) Endurance of device. Switching between the erase state (red triangle) and the program state (blue triangle) is realized by changing the sign of $V_{cg}$ pulse. Pulse amplitude is 30 V and pulse width is 300 ms. A high current on/off ratio (over $10^7$) is observed. The Inset shows that current decay is negligible after 2000 cycles.

**Figure 3. Positive photoconductance and Negative photoconductance observed in ReS$_2$/h-BN/MoS$_2$ heterostructure devices.** (a) Gate-tunable transition between the PPC and the NPC. The thickness of h-BN is 17.4 nm. The device is exposed to 520nm light (73μW). The PPC (blue curve) and the NPC (red curve) could be observed under the control gate pulse of +60V and -60V, respectively. (b) Time-dependent current curves with light illumination of 520 nm (331μW). Blue curve represents photocurrent for ReS$_2$/h-BN/MoS$_2$ and red curve is photocurrent for ReS$_2$/h-BN. The thickness of h-BN is 11.6 nm. The Inset is an optical image of the device, the scale bar is 10 μm. (c) The NPC in MoS$_2$/h-BN/ReS$_2$ device under different control gate voltages. (d) Band diagrams of ReS$_2$/h-BN/MoS$_2$ device before (the upper panel) and after (the lower panel) light illumination.

**Figure 4. Light intensity-controlled NPC effect and NPC-based multi-bit response.** (a) Time-dependent current curves under different intensities of 637nm wavelength. (b) Time-dependent current curves under different light intensities of 520 nm wavelength. Light is switched on and kept for 10 s, and then turned off. '0','1','2' and '3' are labelled as different readout current levels, representing minimum current corresponding to each light intensity (c) The readout current as a function of state ('0', '1','2','3').

**Figure 1**

a) 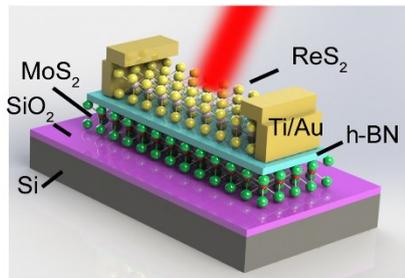

b) 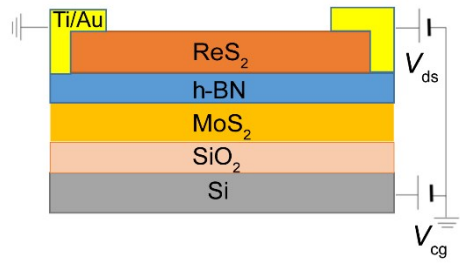

c) 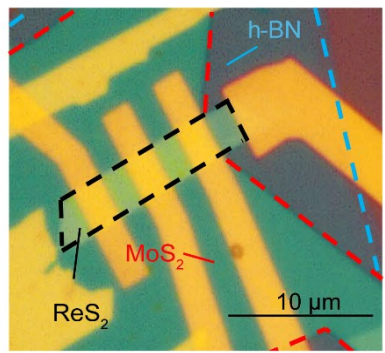

d) 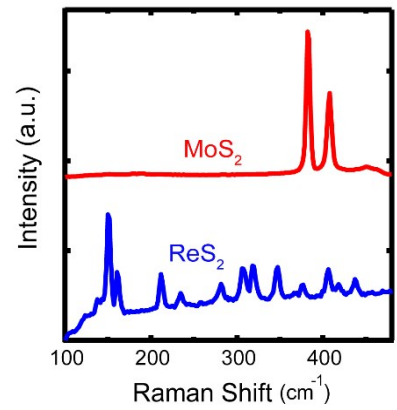

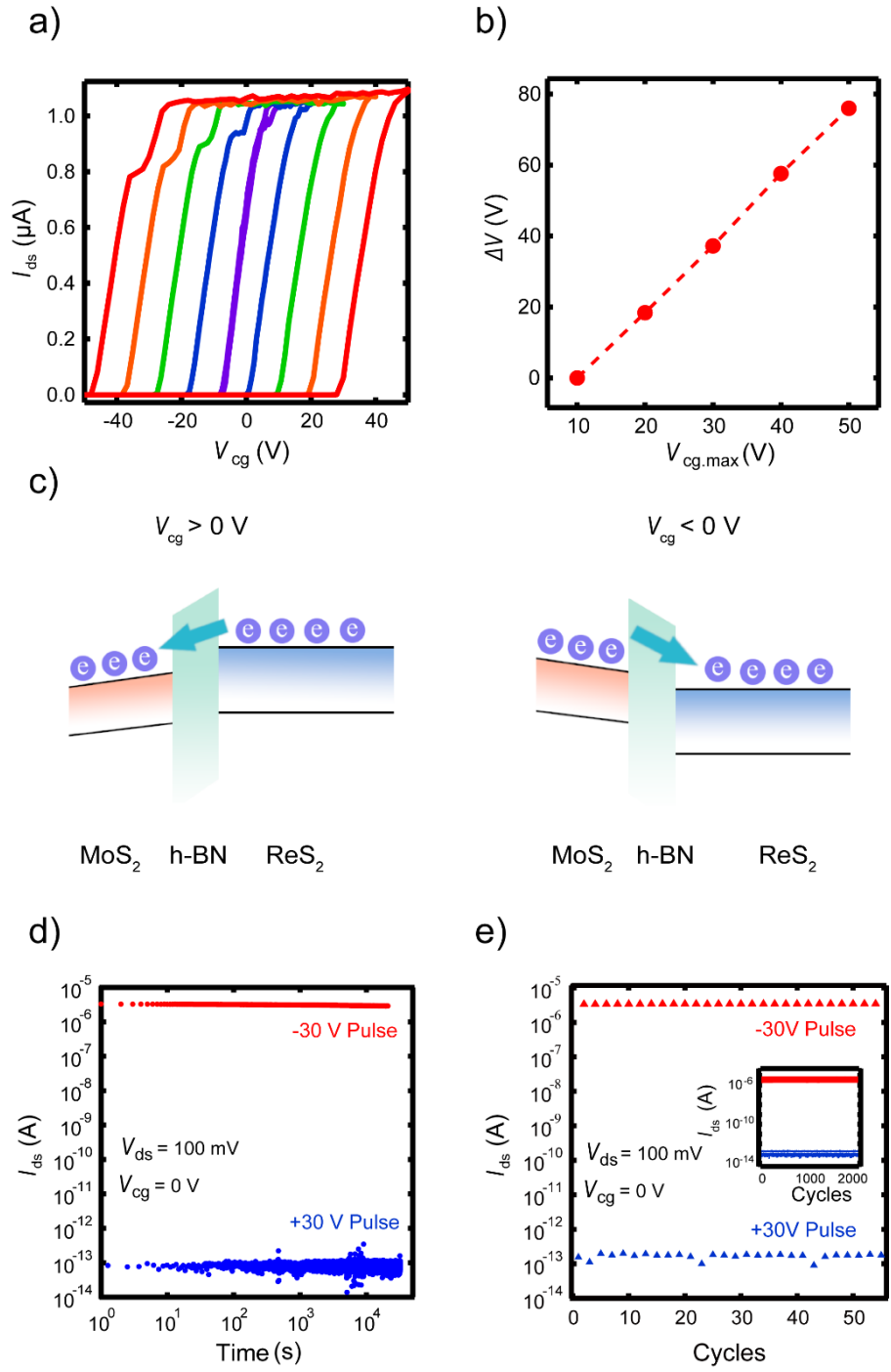

**Figure 3**

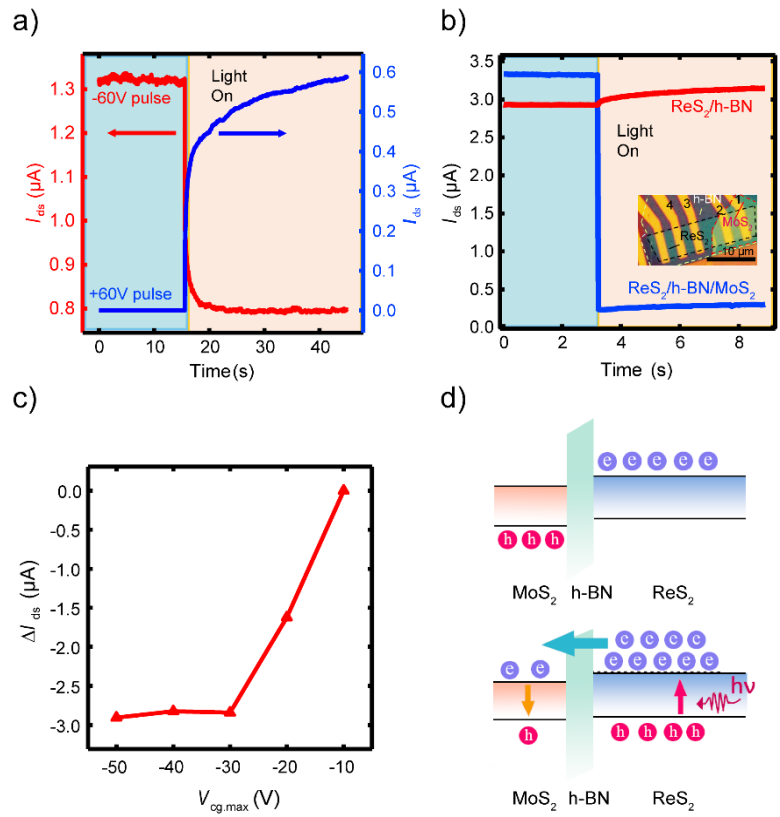

**Figure 4**

a)
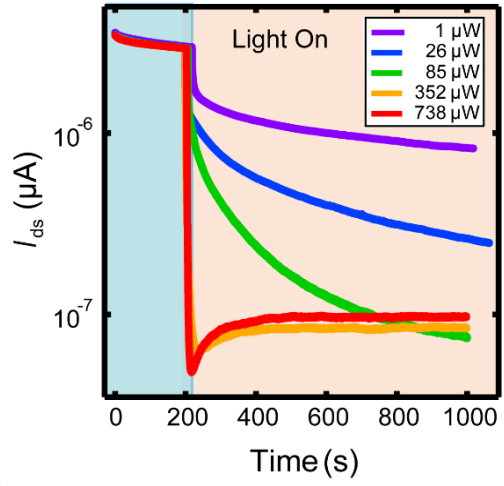

b)
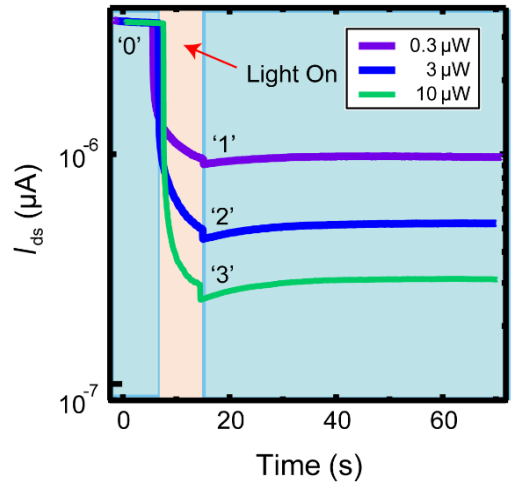

c)
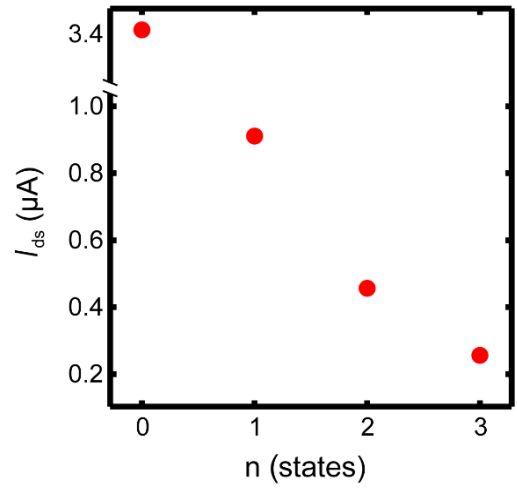